\documentclass[11pt]{article}

\usepackage{amssymb,amsmath,amsfonts,amsthm,lscape}

\graphicspath{{images/}} % cartella dove sono riposte le immagini
 \usepackage{graphicx}

\begin{document}

\thispagestyle{empty}

\

 \vskip1cm

\noindent {\Large{\bf {The classical Darboux III oscillator: factorization, Spectrum\\[6pt] Generating Algebra and solution to the equations of motion
}}}

\medskip
\medskip
\medskip
\medskip

\begin{center}
{\sc \'Angel Ballesteros$^a$, Alberto Enciso$^b$, Francisco J. Herranz$^{a}$,
 \\[4pt] Danilo Latini$^{c}$,  Orlando Ragnisco$^{c,}$\footnote{
 Based on the contribution presented at ``The XXIII International Conference on Integrable Systems and Quantum Symmetries" (ISQS-23),
June 23-27, 2015, 
 Prague,  Czech Republic.} and Danilo Riglioni$^{c}$}
 \end{center}

\medskip
\medskip

\noindent
{$^a$ Departamento de F{\'{\i}}sica, Universidad de Burgos, E-09001 Burgos, Spain}

\medskip

\noindent{$^b$ Instituto de Ciencias Matem\'aticas,  CSIC, Nicol\'as Cabrera 13-15, E-28049 Madrid,
Spain}

\medskip

\noindent{$^c$ Dipartimento di   Matematica e Fisica, Universit\`a di Roma Tre and Istituto Nazionale di Fisica Nucleare sezione di Roma Tre, Via Vasca Navale 84, I-00146 Roma, Italy}

\medskip
 
 \medskip

\noindent{E-mail:\quad {\tt angelb@ubu.es, aenciso@icmat.es, fjherranz@ubu.es,    latini@fis.uniroma3.it, ragnisco@fis.uniroma3.it, riglioni@fis.uniroma3.it}}

\medskip

\medskip

 \medskip

 \medskip

%=========Abstract================
\begin{abstract}
\noindent In a recent paper  the so-called {\em Spectrum Generating Algebra} (SGA) technique has been applied to the $N$-dimensional Taub-NUT system, a maximally superintegrable Hamiltonian system which can be interpreted as a  one-parameter deformation of the Kepler--Coulomb system. Such a Hamiltonian  is  associated to a specific Bertrand space  of non-constant curvature. The SGA procedure unveils the symmetry algebra underlying the Hamiltonian system and, moreover, enables one to solve the equations of motion. Here we will follow the same path to tackle the Darboux III system, another  maximally superintegrable system, which can indeed be viewed as a natural deformation of the isotropic harmonic oscillator  where the flat Euclidean space is again replaced by another space of non-constant curvature.
\end{abstract}

\newpage

%%%%%%%%%%%%%%%%%%%%%%%%%%%%%%%%%%%%%%%%%%%%%%%%%%%

\section{Introduction}

 We consider the classical Hamiltonian in $\mathbb{R}^N$ given by:
\begin{equation}
\label{hclassica}
\mathcal{H}_{\lambda} ({\mathbf{q}}, {\mathbf{p}}) = \mathcal{T}_{\lambda} ({\mathbf{q}}, {\mathbf{p}}) + \mathcal{U}_{\lambda}({\mathbf{q}}) = \frac{{\mathbf{p}}^2}{2m(1 + \lambda{\mathbf{q}}^2)} + \frac{m \omega^2  {\mathbf{q}}^2}{2(1 + \lambda{\mathbf{q}}^2)}  ,
\end{equation}
where $\omega$ and  $\lambda$ are real positive parameters, ${\mathbf{q}}=(q_1,\dots,q_N)$, ${\mathbf{p}}=(p_1,\dots,p_N)$ $\in \mathbb{R}^N$ are conjugate coordinates and momenta, ${\mathbf{q}}^2=\sum_{i=1}^Nq_i^2$ and ${\mathbf{p}}^2=\sum_{i=1}^Np_i^2$.
The kinetic energy $\mathcal{T}_{\lambda} ({\mathbf{q}}, {\mathbf{p}})$ can be interpreted as the
 Lagrangian generating the geodesic motion of a particle of mass $m$ on a conformally flat space known as the Darboux space of type III (D-III). As remarked in \cite{AP2011}, such an $N$-dimensional ($N$D)  curved space is the  spherically symmetric generalization of the D-III surface~\cite{Konig, JMP2003, Pogosyan}, which was constructed in \cite{PLB, PhysicaD}. Moreover, the central potential $\mathcal{U}_{\lambda}({\mathbf{q}})$ was proven to be an `intrinsic oscillator' potential on that space \cite{PhysicaD, AP2009}. We recall that this definition was first proposed in \cite{PhysicaD} and that a generalization of the Bertrand's Theorem \cite{Bertrand} to 3D conformally flat Riemannian spaces has been developed in~\cite{Perlick, CQG2008, CMP2009}. In this respect, we shall say that the Hamiltonian (\ref{hclassica}) determines the D-III oscillator.

We also notice that  in~\cite{PhysicaD} the Hamiltonian $\mathcal{H}_{\lambda}$ has been proven to be maximally superintegrable  by taking advantage of its super-separability (like the usual harmonic oscillator, the system turns out to be separable both in  Cartesian and (hyper)spherical coordinates). This property has been further  analysed in a number of subsequent papers (see, for instance,~\cite{CMP2009, PLA, IJP2011} and references therein), where its algebraic content was explained in terms of the    \emph{Demkov--Fradkin tensor} \cite{Demkov, AJ1965} given by 
 \begin{equation}
\mathcal{ I}_{ij} = p_ip_j+m^2\biggl( \omega^2 -\frac {2\lambda}{m} \mathcal{H}_{\lambda}(\mathbf{q},\mathbf{p}) \biggl) q_iq_j \, , \qquad i,j=1,\dots, N\, .
\label{fradkin}\nonumber
\end{equation}
We recall here that the Hamiltonian $\mathcal{H}_\lambda$  can be expressed in terms of hyperspherical coordinates $r$, $\theta_j$ and canonical momenta $p_r$, $p_{\theta_j}$ , ($j = 1$, $\dots$, $N-1$) defined by  

\begin{equation}
q_j=r \cos \theta_j  \prod_{k=1}^{j-1} \sin \theta_k \quad (1 \le j \le N-1) \, , \qquad q_N=r  \prod_{k=1}^{N-1} \sin \theta_k \, ,
\label{eq:hyper}\nonumber
\end{equation}
such that
\begin{equation}
r=|\mathbf{q}| \, , \qquad \mathbf{p}^2=p_r^2+\frac{\mathbf{L}^2}{r^2}\, , \quad \text{with}\quad \mathbf{L}^2=\sum_{j=1}^{N-1}p_{\theta_j}^2  \prod_{k=1}^{j-1} \frac{1}{\sin^2 \theta_k} \, ,
\label{hyp}\nonumber
\end{equation}
where the radial coordinate $r$ is canonically conjugated to the radial momentum $p_r$. Then, for a fixed value of $\mathbf{L}^2=l^2$, the Hamiltonian \eqref{hclassica} can be written as a one-degree of freedom radial system.

  As a matter of fact the system associated with $\mathcal{H}_{\lambda}$ can be considered as a genuine (maximally superintegrable) $\lambda$-deformation of the usual $N$D isotropic harmonic oscillator with frequency $\omega$, since the limit $\lambda \to 0$ yields
\begin{equation}
\mathcal{H}_0 ({\mathbf{p}}, {\mathbf{q}}) =  \frac{{\mathbf{p}}^2}{2m} + \frac{m \omega^2  {\mathbf{q}}^2}{2} \, .
\label{undef}\nonumber
\end{equation}

  We remind also that  both the  classical and the quantum problems have been investigated (see~\cite{PhysicaD, PLA, IJP2011})  in the special case characterized by positive values of the deformation parameter $\lambda$. Also, an in-depth discussion of the $\lambda<0$ case for the classical system can be found in  \cite{IJP2011} where, however, the explicit formula of the trajectory has not been derived.

  In the present work our scope is rather different from the one pursued   in \cite{PhysicaD}. On one hand, we deal just  with the classical system and restrict  our considerations to the  3D case, with  emphasis on the algebraic side of the problem. In this context we show that  the so-called  {\em Spectrum Generating Algebra} (SGA) technique \cite{AP2008, JPCS2012}  provides us with all the necessary ingredients to achieve its solution, \emph{at least in the $\lambda > 0$ case}, in full analogy with the classical Taub-NUT system~\cite{CTN}. 
In particular, we will derive the solution   for both bounded and   unbounded motions.
On the other hand, through a direct approach, we will be able to handle the $\lambda<0$ case as well, getting an explicit formula  for $t=t(r)$ holding in the \emph{punctured open ball} $$0<r<r_s \doteq \frac{1}{\sqrt{|\lambda|}} ,$$and thus making some progress with respect to what had been already derived in the literature~\cite{PhysicaD, PLA, IJP2011}.

The paper is organized as follows: 

\begin{itemize}
\item
In section $2$ we perform the factorization of the classical Hamiltonian and derive the appropriate Poisson algebra for the ordinary isotropic harmonic oscillator, recovering the well-known solution to the equation of motion.

\item 
In section $3$, paraphrasing the procedure followed in \cite{CTN} we solve the equations of motion in the $\lambda >0$ case. As it happened for the Taub-NUT system, the solution is obtained in the form $t= t(r)$. As already noticed in \cite{PhysicaD},  in contrast with the usual harmonic oscillator, for sufficiently high energy values there are unbounded orbits (the particle can escape to infinity) which in the quantum setting corresponds to the onset of a continuous spectrum. Coherently with the predictions of the Bertrand's Theorem, the bounded orbits  will be closed and there will be stable circular trajectories.

\item 
In section $4$ the $\lambda <0$ case will be tersely investigated. We will get behaviours  that  are similar to those exhibited by the Taub-NUT system for negative values of the deformation parameter $\eta$ \cite{CTN}. Indeed, in a sort of dual way with respect  to the Taub-NUT case,  the kinetic energy (and thus the metric) remains positive  in the bounded region $0 <r <r_s$ while it takes negative values for $r>r_s$.  So, in order to get an admissible metric for $r>r_s$, we should  there resort  to time-reversal  symmetry to restore positivity. On the other hand, in  that unbounded region  the effective potential is a monotonic function  of $r$ having a constant limiting value at  infinity, implying that no  closed orbits can arise.

\item 
The main results and some further perspectives are outlined in the concluding section. 

\end{itemize}

%%%%%%%%%%%%%%%%%%%%%%%%%%%%%%%%%%%%%%%%%%%%%%%%%%%

\section{The Euclidean case: isotropic harmonic oscillator}

 Before using the machinery of the SGA  approach to tackle the $\lambda \neq 0$ case, we will firstly characterize the Euclidean case in order to check our future results in the (flat) limit $\lambda \to 0$ and to illustrate the usual SGA procedure.  Therefore,  let us consider the Hamiltonian
\begin{equation}
H_0(r,p)=T_0(r, p)+V_{{\rm 0,  eff}}(r)=\frac{p^2}{2m}+\frac{l^2}{2m r^2}+\frac{1}{2}m\omega^2 r^2 \, ,
\label{hamosc}
\end{equation}
where $r$ is the radial coordinate and $p\equiv p_r$ is the radial momentum related to the particle of mass $m$, frequency $\omega$ and angular momentum $l$.   The effective potential is so given by

\begin{equation}
V_{{\rm 0,  eff}}(r)=\frac{l^2}{2m r^2}+\frac{1}{2}m\omega^2 r^2 \, ,
\label{effpot}
\end{equation}

\begin{figure}
\centering
\includegraphics[width=7.5cm, height=5.5cm]{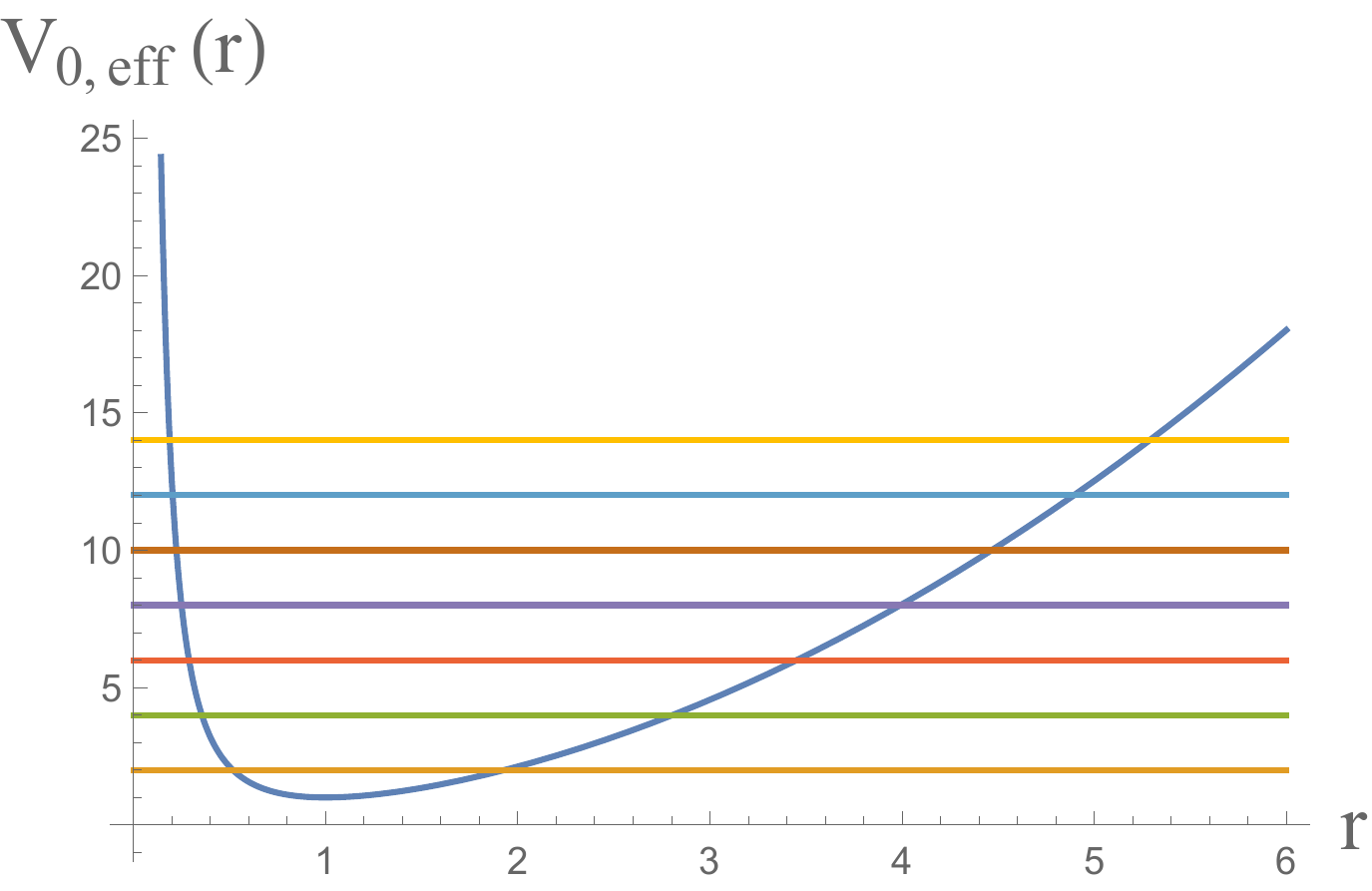}
\quad
\includegraphics[width=7.5cm, height=5.5cm]{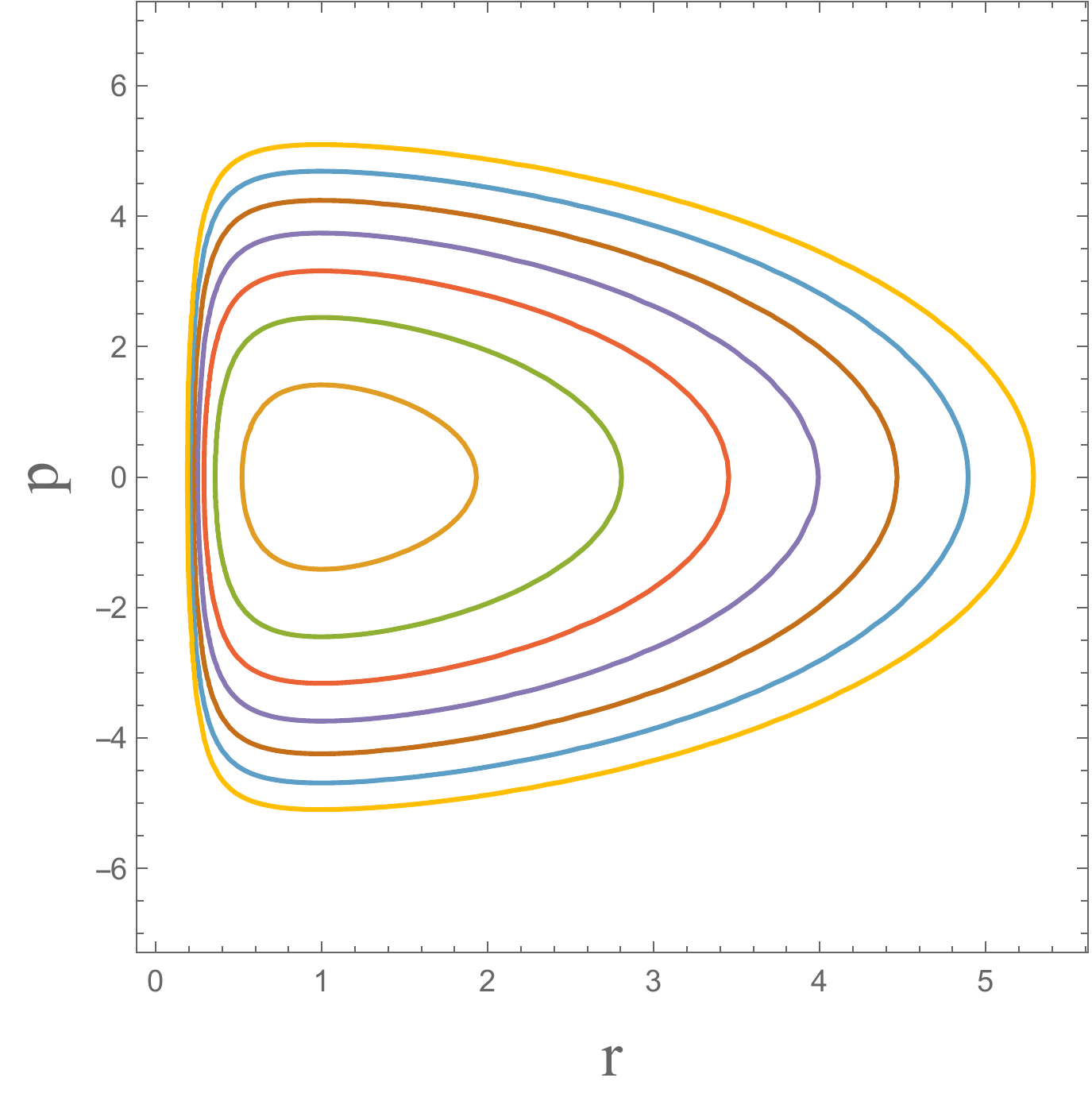}
\caption{Effective potential $V_{\rm 0, eff}(r)$ (\ref{effpot}) and phase plane $(r, p)$ calculated for $m=\omega=l=1$ and $E=2, 4, 6, 8, 10, 12,14$ (in appropriate units).}
\label{fig1}
\end{figure}

Clearly, in order to have closed orbits, the values of the energy have to be confined in the region $V_{ {\rm 0,  eff}}(r_{\rm min})<E<\infty$, where $V_{ {\rm 0,  eff}}(r_{\rm min})= l \omega $ is the minimum of the effective potential, i.e. its value calculated at $r_{\rm min}=\sqrt{\frac{l}{m \omega}}$ (see lhs of  figure \ref{fig1}).

 Now, in order to apply the SGA technique, we proceed by factorizing the Hamiltonian \eqref{hamosc} as follows
\begin{equation}
p^2 r^2+m^2 \omega^2 r^4-2 m r^2 H_0=A_0^+A_0^-+\gamma(H_0)=-l^2 \, ,
\label{factoriz}
\end{equation}
where $A_0^{\pm}$ are unknown functions to be determined. In particular, taking into account  \eqref{factoriz}, we obtain
\begin{equation}
A_0^{\pm} = \mp i r p + m \omega r^2 -\frac{H_0}{\omega}  \, ,
\label{eq:ansatz}
\end{equation}
and then 
\begin{equation}
\gamma(H_0)=-\frac{H_0^2}{\omega^2} . \label{za}
\end{equation}
 The Hamiltonian $H_0$ and $A_0^{\pm}$ are functions of the canonical coordinates $(r,p)$ and they have to close the following Poisson algebra \cite{AP2008, JPCS2012}:
\begin{align}
\{H_0,A_0^{\pm} \}=\mp i \alpha(H_0) A_0^{\pm} \, ,\qquad 
\{A_0^+,A_0^-\}= i \beta(H_0) \, \nonumber .
\label{eq:poissoan}
\end{align}
In particular, using the expressions \eqref{eq:ansatz}, we obtain that
\begin{equation}
\alpha(H_0)=2   \omega \,, \qquad \beta(H_0)=\frac {4 H_0}{\omega} \, .
\label{poissonAaA}\nonumber
\end{equation}
 Summarizing, we have the following relations 
\begin{align}
\{H_0,A_0^{\pm} \}=\mp \, i 2  \omega  A_0^{\pm}\, , 
\qquad  \{A_0^+,A_0^- \}= i \frac{4 H_0}{\omega} \, .
\label{poi}
\end{align}
At this point we can define the \emph{time-dependent constants of motion}
\begin{equation}
Q_0^{\pm}=A_0^{\pm} e^{\mp i \alpha(H_0) t} \, ,
\label{eq:constants}\nonumber
\end{equation}
such that 
$$
\frac{{\rm d} Q_0^{\pm}}{{\rm d}t} =\{Q_0^{\pm},H_0 \}+ \frac{\partial Q_0^{\pm}}{ \partial t} =0 \, .
$$
As we know, those dynamical  variables can be written as $Q_0^{\pm}=q_0 e^{\pm i \theta_0}$ and allow us to determine the motion \cite{AP2008, JPCS2012, CTN}. Indeed it holds:
 \begin{equation}
 \left(\mp i r p + m \omega r^2 -\frac{H_0}{\omega} \right)\, e^{\mp 2 i   \, \omega  t}=q_0e^{  \pm i \theta_0} \, .
\label{motocons}\nonumber
\end{equation}
By imposing (\ref{factoriz})  along (\ref{za}),
that is,
$$
A_0^+A_0^- -\frac{H_0^2}{\omega^2}=q_0^2-\frac{H_0^2}{\omega^2}=-l^2 \, ,
$$
we find that
\begin{align}
\begin{cases}
& - i r p + m \omega r^2 -\dfrac{H_0}{\omega}  =q_0e^{ i \bigl(2 \, \omega  t + \theta_0\bigl)} \, ,\\
&  i r p + m \omega r^2 -\dfrac{H_0}{\omega}  =q_0e^{- i\bigl(2 \, \omega  t + \theta_0\bigl)} \, ,
\label{eq:moto}
\end{cases}
\end{align}
where 
$$
q_0=\sqrt{-l^2+\frac{H_0^2}{\omega^2}} \, .
$$

%%%%%%%%%%%%%%%%%%%%%%%%%%%%%%%%%%%%%%%%%%

%%%%%%%%%%%%%%%%%%%%%%%%%%%%%%%%%%%%%%%%%%%%%%%%%%%

Now, since the Hamiltonian does not depend explicitly on time, it is a constant of motion, i.e. the energy of the system $H_0=E$. Therefore, summing and subtracting \eqref{eq:moto} we obtain (on the level surface $H_0=E$)
\begin{align}
\begin{cases}
& m \omega r^2 -\dfrac{E}{ \omega}= q_0 \cos \left(2  \, \omega  t + \theta_0\right) \, ,\\
&rp=-q_0 \sin \left(2  \, \omega  t + \theta_0\right)\, .
\label{eq:poi}
\end{cases}
\end{align}
Finally, using \eqref{eq:poi} we are able to write $r$ and $p$ as  functions of $t$:
\begin{align}
\begin{cases}
&r(t)=  \sqrt{\dfrac{ E}{m \omega^2}+\dfrac{q_0}{m \omega} \cos \left(2 \, \omega  t + \theta_0\right)} \, ,\\[10pt]
&p(t)=- \dfrac{q_0 \sin \left(2  \, \omega  t + \theta_0\right)}{\sqrt{\dfrac{ E}{m \omega^2}+\dfrac{q_0}{m \omega} \cos \left(2 \, \omega  t + \theta_0\right)}} \, .
\label{eq:conti}
\end{cases}
\end{align}

%%%%%%%%%%%%%%%%%%%%%%%%%%%%%%%%%%%%%%%%%%%%%%%%%%%%
\begin{figure}
\centering
\includegraphics[width=7.5cm, height=5.5cm]{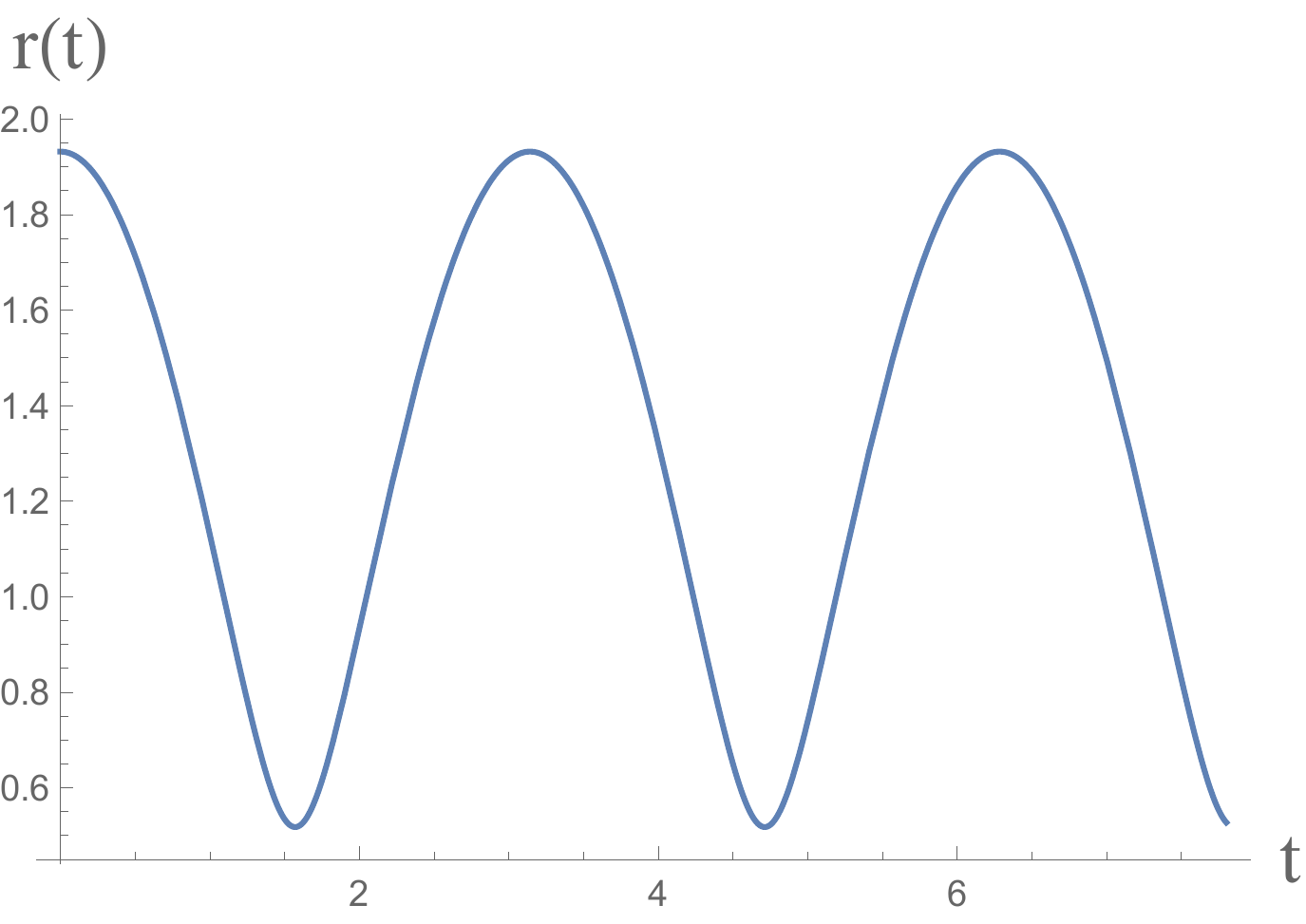}
\quad
\includegraphics[width=7.5cm, height=5.5cm]{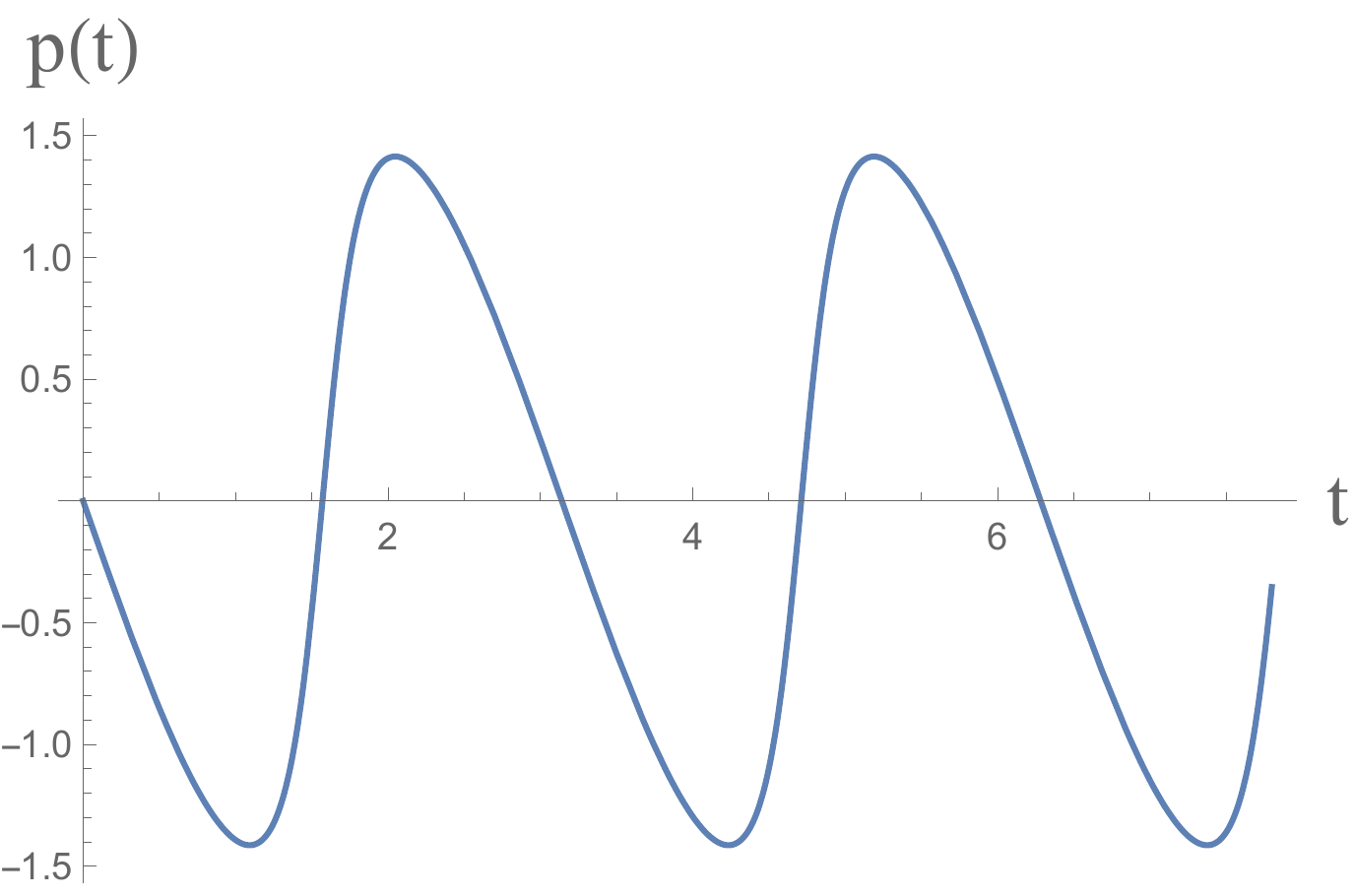}
\caption{Trajectory $r(t)$ and momentum $p(t)$ calculated for $m=\omega=l=1$ and $E=2$.}
\label{fig2}
\end{figure}
%%%%%%%%%%%%%%%%%%%%%%%%%%%%%%%%%%%%%%%%%%%%%%%%%%%%

 In this way we have found $r(t)$ and $p(t)$ and then the motion has been fully determined by means of the SGA procedure.
 In the rhs of figure~\ref{fig1} the phase plane  is depicted, while in figure \ref{fig2} the curves (\ref{eq:conti}) are represented.

  In the following, we shall deal with the deformed case using the same procedure. Clearly, the results just obtained in this section have to be recovered in the Euclidean (i.e. $\lambda \to 0$) limit.

%%%%%%%%%%%%%%%%%%%%%%%%%%%%%%%%%%%%%%%%%%%%%%%%%%%

\section{The deformed $\lambda>0$ case: D-III oscillator}

\begin{figure}
\centering
\includegraphics[width=7.5cm, height=5.5cm]{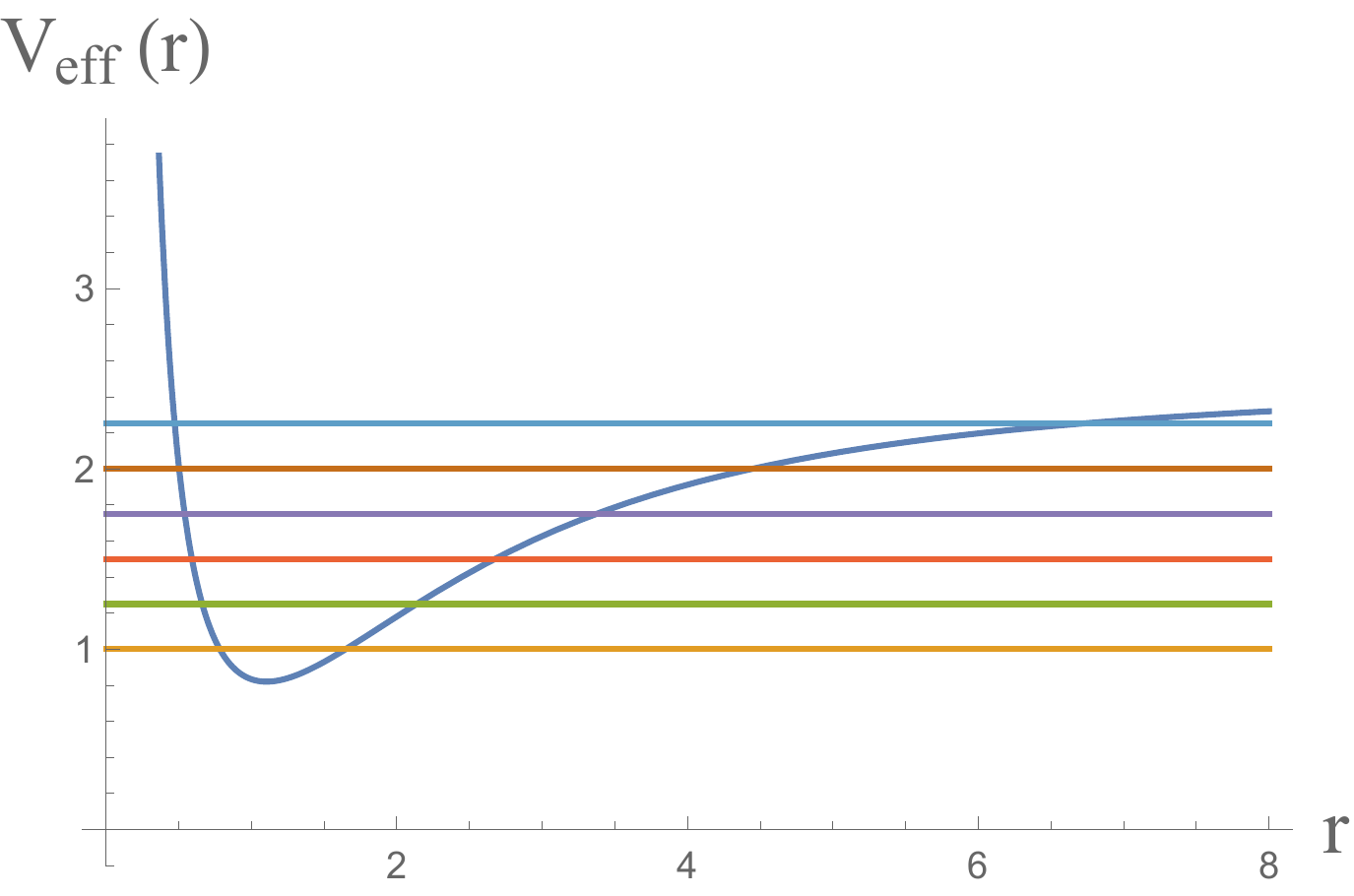}
\quad
\includegraphics[width=7.5cm, height=5.5cm]{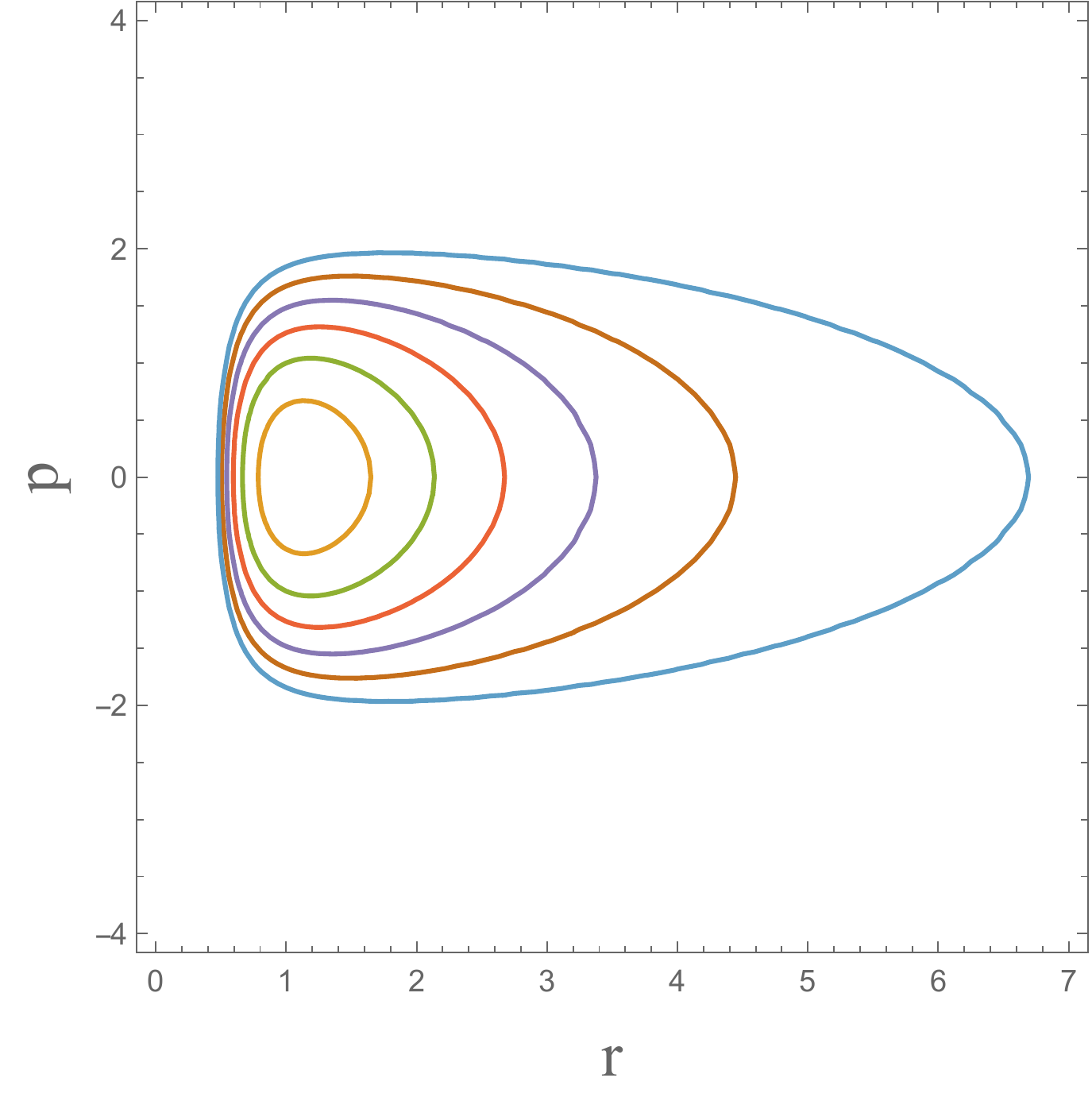}
\caption{Effective potential $V_{\rm eff}(r)$ (\ref{hamm}) and phase plane $(r, p)$ calculated for $m=\omega=l=1$ and $E=1.00, 1.25, 1.50, 1.75, 2.00, 2.25$. The deformation parameter is fixed at the value $\lambda=0.20$.}
\label{fig3}
\end{figure}

\begin{figure}
\centering
\includegraphics[width=7.5cm, height=5.5cm]{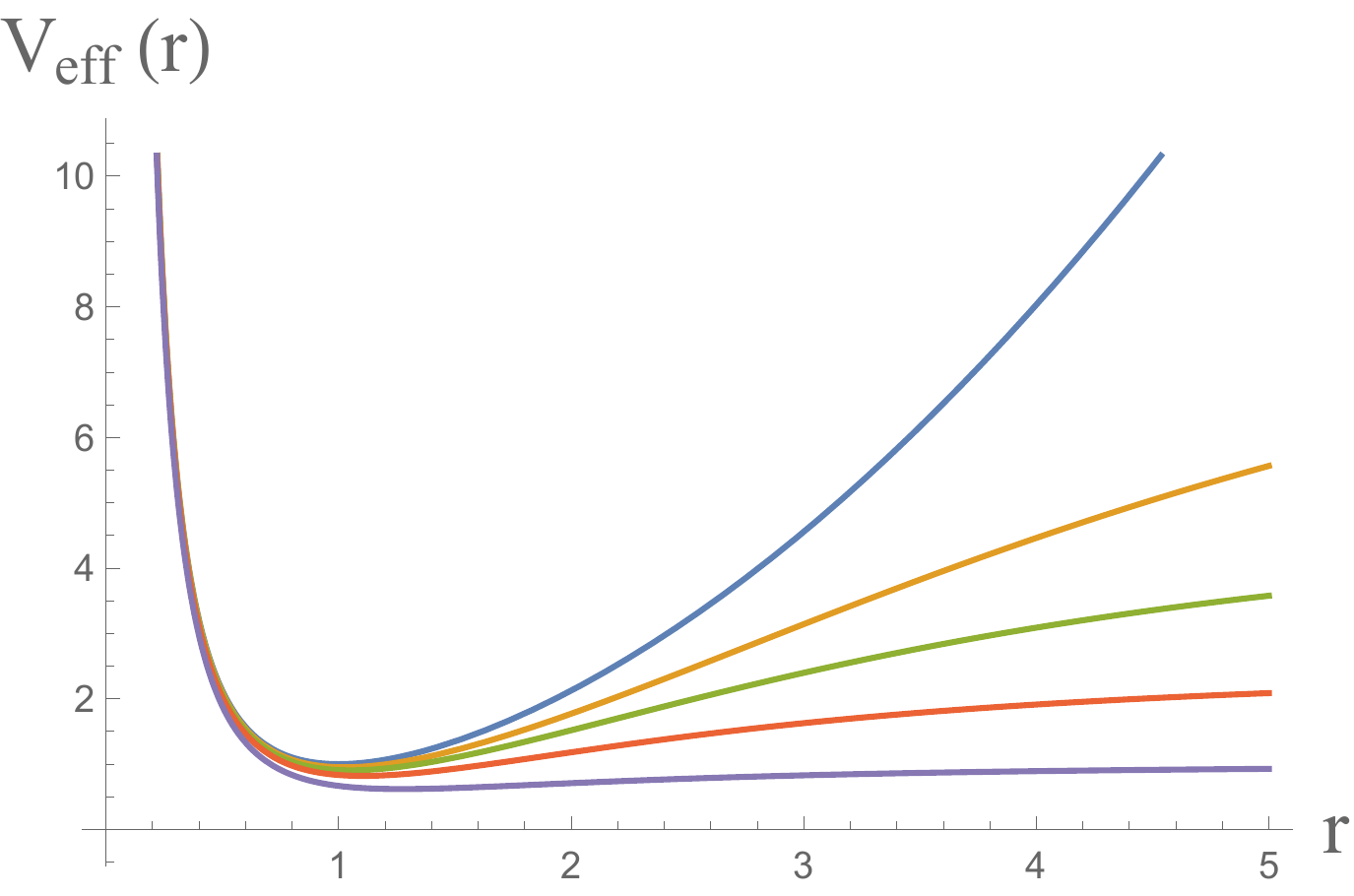}
\quad
\includegraphics[width=7.5cm, height=5.5cm]{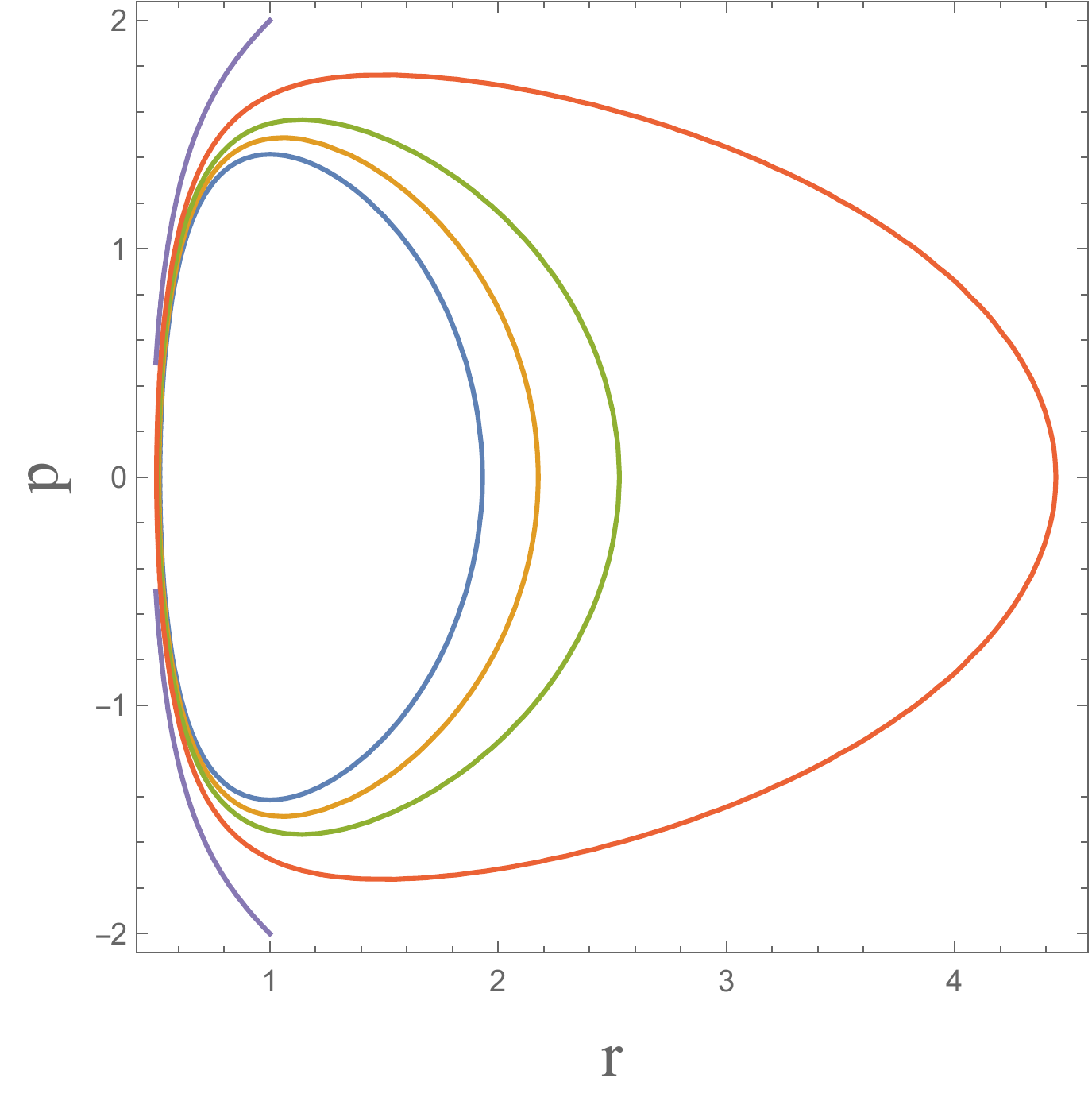}
\caption{Effective potential $V_{\rm eff}(r)$  (\ref{hamm})  and phase plane $(r, p)$ calculated for $m=\omega=l=1$ and $E=2$. The deformation parameter take the values $\lambda=0, 0.05, 0.10, 0.20, 0.50$.}
\label{fig3a}
\end{figure}

 In this section we focus our attention on the D-III oscillator Hamiltonian (\ref{hclassica}) written as a 1D radial system by means of hyperspherical coordinates, namely
\begin{equation} 
H(r,p)=T(r,p)+V_{\rm eff}(r)=\frac{ p^2}{2m(1+\lambda r^2)}+\frac{ l^2}{2m r^2(1+\lambda r^2)}+\frac{m \omega^2 r^2}{2(1+\lambda r^2)} = \mathcal{F}(r) H_0 \,,
\label{hamm}
\end{equation} 
where $m$, $\omega$ and $l$ are positive constants, $\lambda$ is the deformation parameter, $H_0$ is the `undeformed' isotropic oscillator Hamiltonian \eqref{hamosc} and 
$$
\mathcal{F}(r) \doteq \frac{1}{1+\lambda r^2}\, .
$$
 Multiplying both sides of \eqref{hamm} by $r^2(1+\lambda r^2)$ we get:
\begin{equation}
r^2(1+\lambda r^2)H=r^2\biggl(\frac{ p^2}{2m}+\frac{l^2}{2mr^2}+\frac{1}{2}m \omega^2 r^2 \biggl)=\frac{1}{2m}(r^2p^2+l^2+m^2 \omega^2 r^4) .
\label{fact}
\end{equation}
 Now, as it has been done  in the previous section for the undeformed case, at any $r$ we can  factorize \eqref{fact} as:
\begin{equation}
r^2p^2+ m^2 r^4 \left(\omega^2-\frac{2 \lambda}{m} H\right)-2 m r^2H=A^+A^-+\gamma(H)=-l^2 \, ,
\label{factor}
\end{equation}
 where $A^+$, $A^-$ are unknown functions of  $r$, $p$. We make the following \emph{ansatz} for $A^+$, $A^-$:
\begin{equation}
A^{\pm} = \left(\mp i r p + m r^2 \sqrt{\omega^2-\frac{2 \lambda}{m} H}  -\frac{H}{\sqrt{\omega^2-\frac{2 \lambda}{m} H}} \right)e^{\pm f(r,p)} \, .
\label{eq:ansatzs}\nonumber
 \end{equation}
 The `arbitrary function' $f(r,p)$ will be determined by requiring the closure of the Poisson algebra generated
by $H$ and $A^{\pm}$:
\begin{align}
& \{H,A^{\pm} \}=\mp i \alpha(H) A^{\pm} \, , \nonumber\\
&\{A^+,A^-\}= i \beta(H) \, ,
\label{eq:poisson}\nonumber
\end{align}
 where  the functions $\alpha$, $\beta$ have to be determined. Inserting $A^{\pm}$ in \eqref{factor} we get
$$
\gamma(H)=-\frac{H^2}{\omega^2-\frac{2 \lambda}{m} H} \, ,
$$
 and requiring that  $A^{\pm}$ obey the proper Poisson brackets we arrive at
\begin{equation}
f(r,p)=- \frac{i \lambda r p \sqrt{\omega^2-\frac{2 \lambda}{m} H}}{m (\omega^2-\frac{\lambda}{m} H)}\,\, , \qquad \alpha(H) = \frac{2 \bigl( \omega^2-\frac{2 \lambda}{m} H \bigl)^{\frac{3}{2}}}{\omega^2-\frac{\lambda}{m} H} \,\, , \qquad \beta(H)=\frac{4  H}{\sqrt{\omega^2-\frac{2 \lambda}{m}H}} \, .
\label{poissonAA}\nonumber
\end{equation}
 Hence  we find that
\begin{equation}
A^{\pm}= \biggl(\mp i r p + m r^2 \sqrt{\omega^2-\frac{2 \lambda}{m} H}  -\frac{H}{\sqrt{\omega^2-\frac{2 \lambda}{m} H}} \biggl)\exp  \left\{ {\mp  \frac{i \lambda r p \sqrt{\omega^2-\frac{2 \lambda}{m} H}}{m (\omega^2-\frac{\lambda}{m} H)}} \right\} ,
\label{ladder}\nonumber
\end{equation}
\begin{equation}
\{H,A^{\pm} \}=\mp i \,  \frac{2 \bigl( \omega^2-\frac{2 \lambda}{m} H \bigl)^{\frac{3}{2}}}{\omega^2-\frac{\lambda}{m} H} \, A^{\pm} \,\, , \qquad  \{A^+,A^- \}=i \frac{4  H}{\sqrt{\omega^2-\frac{2 \lambda}{m}H}} \, .
\label{poissonbrackets}\nonumber
\end{equation}
  We notice that in the limit $\lambda \to 0$  one gets back the undeformed Poisson algebra \eqref{poi} as expected. We also point out that the requirement $\omega^2-\frac{2 \lambda}{m} H>0$ implies the upper bound $E<\frac{m \omega^2}{2 \lambda}$. Moreover, for bounded motion, the energy has to be greater than the minimum of the effective potential $V_{\rm eff}(r)$. The latter turns out to be 
  $$
  V_{\rm eff}(r_{\rm min})=\frac{l^2}{m}\left(\sqrt{\lambda^2+\frac{m^2\omega^2}{ l^2}}-\lambda \right) \, ,
  \qquad  r^2_{\rm min}=\frac{l^2}{m^2\omega^2}\left(\lambda+\sqrt{\lambda^2+\frac{m^2 \omega^2}{l^2}} \, \right) \, .
  $$
As a consequence the energy belongs to the interval    (see lhs of figures \ref{fig3} and  \ref{fig3a})
$$
V_{\rm eff}(r_{\rm min})<E<\frac{m \omega^2}{2 \lambda} \, .
$$

 Next, as usual, in order to find the corresponding trajectory we define the \emph{time-dependent constants of the motion}
\begin{equation}
Q^{\pm}=A^{\pm} e^{\mp i \alpha(H) t} \, ,
\nonumber
\end{equation}
from which we can write (on-shell):
\begin{align}
& \left(\mp i r p + m r^2 \sqrt{\omega^2-\frac{2 \lambda}{m} E}  -\frac{E}{\sqrt{\omega^2-\frac{2 \lambda}{m} E}} \right)
  \exp \left\{ {\mp  i \left(\frac{\lambda r p \sqrt{\omega^2-\frac{2 \lambda}{m} E}}{m (\omega^2-\frac{\lambda}{m} E)}+\frac{2 \bigl( \omega^2-\frac{2 \lambda}{m} E \bigl)^{\frac{3}{2}}}{\omega^2-\frac{\lambda}{m} E}\, t \right)} \right\}
  \cr
  &\qquad\qquad =q\, e^{  \pm i \theta} \, ,
\label{moto}\nonumber
\end{align}
or else
\begin{equation}
\begin{cases}
- i r p + m r^2 \sqrt{\omega^2-\frac{2 \lambda}{m} E}  -\dfrac{E}{\sqrt{\omega^2-\frac{2 \lambda}{m} E}} =q  \exp\left\{ { i\biggl(\frac{\lambda r p \sqrt{\omega^2-\frac{2 \lambda}{m} E}}{m (\omega^2-\frac{\lambda}{m} E)}+\frac{2 \bigl( \omega^2-\frac{2 \lambda}{m} E \bigl)^{\frac{3}{2}}}{\omega^2-\frac{\lambda}{m} E } \, t  +\theta \biggl)} \right\} \, ,\\
+ i r p + m r^2 \sqrt{\omega^2-\frac{2 \lambda}{m} E}  -\dfrac{E}{\sqrt{\omega^2-\frac{2 \lambda}{m} E}}=q \exp\left\{ {- i \biggl(\frac{\lambda r p \sqrt{\omega^2-\frac{2 \lambda}{m} E}}{m (\omega^2-\frac{\lambda}{m} E)}+\frac{2 \bigl( \omega^2-\frac{2 \lambda}{m} E \bigl)^{\frac{3}{2}} }{\omega^2-\frac{\lambda}{m} E} \, t +\theta \biggl)} \right\}\, ,
\end{cases}
\label{ecoupled}
\end{equation}
where  now 
$$q=\sqrt{-l^2+\frac{E^2}{\omega^2-\frac{2 \lambda}{m} E}} \, .
$$
 Once again, summing and subtracting \eqref{ecoupled} we obtain:
\begin{equation}
\begin{cases}
 m r^2 \sqrt{\omega^2-\dfrac{2 \lambda}{m} E}  -\dfrac{E}{\sqrt{\omega^2-\frac{2 \lambda}{m} E}}=  q \cos \left(\dfrac{\lambda r p \sqrt{\omega^2-\frac{2 \lambda}{m} E}}{m (\omega^2-\frac{\lambda}{m} E)}+\dfrac{2 \bigl( \omega^2-\frac{2 \lambda}{m} E \bigl)^{\frac{3}{2}}}{\omega^2-\frac{\lambda}{m} E } \, t  +\theta  \right) \, ,\\[20pt]
 r p = - q  \sin\left(\dfrac{\lambda r p \sqrt{\omega^2-\frac{2 \lambda}{m} E}}{m (\omega^2-\frac{\lambda}{m} E)}+\dfrac{2 \bigl( \omega^2-\frac{2 \lambda}{m} E \bigl)^{\frac{3}{2}}}{\omega^2-\frac{\lambda}{m} E } \, t  +\theta \right)\, .
\end{cases}
\label{xxx}\nonumber
\end{equation}
Taking the sum of the square of these two equations we recover $\eqref{factor}$ restricted to the level surface $H=E$.
 Finally, thanks to the above relations, we are able to express  $t$ as a  function of $r$:
\begin{align}
& t(r)=\frac{1}{\Omega_{[\lambda]}(E)}\left[\arccos \left(\frac{ m r^2 (\omega^2-\frac{2 \lambda}{m} E ) - E }{ q \sqrt{\omega^2-\frac{2 \lambda}{m} E}}\right)  \right.\cr
&\qquad\qquad \qquad\quad  - \left. \frac{\lambda \sqrt{\omega^2-\frac{2 \lambda}{m} E}}{m(\omega^2-\frac{ \lambda}{m} E)}\sqrt{2 m E r^2-l^2-m^2 r^4 \bigl(\omega^2-\tfrac{2 \lambda}{m} E\bigl)}-\theta \right] \,  ,
\label{eq:traject}
\end{align}
 where 
 $$
 \Omega_{[\lambda]}(E)=\frac{2 \bigl( \omega^2-\frac{2 \lambda}{m} E \bigl)^{\frac{3}{2}}}{\omega^2-\frac{\lambda}{m} E} \equiv \alpha(E)
 $$
  is the angular frequency of the  motion.  In the rhs of figures \ref{fig3} and  \ref{fig3a} some representations of the resulting phase space  can be found.

 %%%%%%%%%%%%%%%%%%%%%%%%%%%%%%%%%%%%%%%%%%%%%%%%%%%%.
%%%%%%%%%%%%%%%%%%%%%%%%%%%%%%%%%%%%%%%%%%%%%%%%%%5

%%%%%%%%%%%%%%%%%%%%%%%%%%%%%%%%%%%%%%%%%%%%%%%%%%

 At this point, defining the quantities 
 $$
 a^2_{[\lambda]} \doteq \frac{E}{m\bigl(\omega^2-\frac{2 \lambda}{m} E \bigl)} \,  ,\qquad\epsilon_{[\lambda]} \doteq \sqrt{1-\frac{\bigl(\omega^2-\frac{2 \lambda}{m} E\bigl) l^2}{E^2} } \, , 
 $$
 $a^2_{[\lambda]}$ being the square of the major semi-axes of the ellipse and  $\epsilon_{[\lambda]}$ a parameter directly related to its eccentricity,  equation \eqref{eq:traject} can be cast into the following form:
\begin{equation}
\Omega_{[\lambda]}(E) t(r)+\theta=\arccos \biggl[-\frac{1}{\epsilon_{[\lambda]}}\biggl(1-\biggl(\frac{r}{a_{[\lambda]}}\biggl)^2 \biggl) \biggl]-\frac{\lambda a^2_{[\lambda]} }{1+\lambda a^2_{[\lambda]}} \sqrt{\epsilon^2_{[\lambda]}-\biggl[1-\biggl(\frac{r}{a_{[\lambda]}}\biggl)^2 \biggl]^2} \, ,
\label{eq:tradefo}
\end{equation}
whose structure is very similar to the one obtained for the classical Taub-NUT  system~\cite{CTN}. 
 In the limit $\lambda \to 0$ the equation \eqref{eq:tradefo} can be easily inverted to obtain the trajectory $r(t)$ given in \eqref{eq:conti}, showing that it just represents the $\lambda$-deformation of the solution of the Euclidean isotropic harmonic oscillator. Then, we can conclude that the motion has been solved by means of the SGA method.

 Now we turn to investigate the unbounded motion for positive values of the deformation parameter (see also~\cite{PhysicaD}). The latter will arise for $E>\frac{m \omega^2}{2 \lambda}$. 
In this case 
$$
\omega^2-\frac{2\lambda}{m}E=-\left(\frac{2\lambda}{m}E-\omega^2\right)<0 \, ,
$$
therefore all the previous functions have to be adapted to this situation. In particular, we can write:
\begin{equation}
\tilde{A}^{\pm}= i \left(\mp  r p + m r^2 \sqrt{\frac{2 \lambda}{m} H-\omega^2}  +\frac{H}{\sqrt{\frac{2 \lambda}{m} H-\omega^2}} \right) \exp\left\{{\pm  \frac{ \lambda r p \sqrt{\frac{2 \lambda}{m} H-\omega^2}}{m (\omega^2-\frac{\lambda}{m} H)}}\right\} \, ,
\label{ladderunb}\nonumber
\end{equation}
\begin{equation}
\tilde{\alpha}(H)= -  i \,\frac{2\bigl(\frac{2 \lambda}{m} H-\omega^2 \bigl)^{\frac{3}{2}}}{\omega^2-\frac{\lambda}{m} H} \,\, , \qquad  \tilde{\beta}(H)= - i\, \frac{4 H}{\sqrt{\frac{2 \lambda}{m} H-\omega^2}} \, .
\label{poissonbracketsunb}\nonumber
\end{equation}
All the functions are purely imaginary and the time-dependent constants of motion (also purely imaginary) take the form (see~\cite{JPCS2012}):
\begin{equation}
\tilde{Q}^{\pm}=\tilde{A}^{\pm}e^{\mp i \tilde{\alpha}(H) t}=i \tilde{q} e^{\mp \theta} \, ,\qquad  \tilde{q}\doteq \sqrt{l^2+\frac{E^2}{\frac{2 \lambda}{m} E-\omega^2}} \, .
\label{comps}\nonumber
\end{equation}
This leads to the two following equations (again on-shell):
\begin{equation}
\begin{cases}
-  r p + m r^2 \sqrt{\frac{2 \lambda}{m} E-\omega^2}  +\dfrac{E}{\sqrt{\frac{2 \lambda}{m} E-\omega^2}} =\tilde{q} \exp\left\{ { -\biggl(\frac{\lambda r p \sqrt{\frac{2 \lambda}{m} E-\omega^2}}{m (\omega^2-\frac{\lambda}{m} E)}-\frac{2 \bigl( \frac{2 \lambda}{m} E-\omega^2 \bigl)^{\frac{3}{2}}}{\omega^2-\frac{\lambda}{m} E } \, t  +\theta \biggl)}  \right\} \,  , \\
+  r p + m r^2 \sqrt{\frac{2 \lambda}{m} E-\omega^2}  +\dfrac{E}{\sqrt{\frac{2 \lambda}{m} E-\omega^2}}=\tilde{q} \exp\left\{ {+ \biggl(\frac{\lambda r p \sqrt{\frac{2 \lambda}{m} E-\omega^2}}{m (\omega^2-\frac{\lambda}{m} E)}-\frac{2 \bigl( \frac{2 \lambda}{m} E-\omega^2 \bigl)^{\frac{3}{2}}}{\omega^2-\frac{\lambda}{m} E} \, t +\theta \biggl)}  \right\}\, ,
\end{cases}
\label{ecoupledunb}\nonumber
\end{equation}
implying that
\begin{equation}
\begin{cases}
 m r^2 \sqrt{\frac{2 \lambda}{m} E-\omega^2}  +\dfrac{E}{\sqrt{\frac{2 \lambda}{m} E-\omega^2}}=  \tilde{q}  \cosh \left(\dfrac{\lambda r p \sqrt{\frac{2 \lambda}{m} E-\omega^2}}{m (\omega^2-\frac{\lambda}{m} E)}-\dfrac{2 \bigl( \frac{2 \lambda}{m} E-\omega^2 \bigl)^{\frac{3}{2}}}{\omega^2-\frac{\lambda}{m} E } \, t  +\theta \right) \, , \\[20pt]
 r p = \tilde{q}  \sinh\left(\dfrac{\lambda r p \sqrt{\frac{2 \lambda}{m} E-\omega^2}}{m (\omega^2-\frac{\lambda}{m} E)}-\dfrac{2 \bigl( \frac{2 \lambda}{m} E-\omega^2 \bigl)^{\frac{3}{2}}}{\omega^2-\frac{\lambda}{m} E } \, t  +\theta \right)\, .
\end{cases}
\label{eqcoupled}\nonumber
\end{equation}
 We notice that the equation \eqref{factor} (for $H=E$) is now recovered by taking the difference of the square of these two equations. Also in this case, due to the above relations, we are able to obtain  $t$ as a  function of $r$, namely:
\begin{align}
& t(r)=\frac{1}{\zeta_{[\lambda]}(E)}\left[\arccos\!\text{h} \left(\frac{m r^2 (\frac{2 \lambda}{m} E-\omega^2) + E}{ \tilde{q}\sqrt{\frac{2 \lambda}{m} E-\omega^2}}\right)  \right.\cr
&\qquad\qquad \qquad\quad  \left. +\frac{\lambda \sqrt{\frac{2 \lambda}{m} E-\omega^2}}{m(\frac{ \lambda}{m} E-\omega^2)}\sqrt{2 m E r^2-l^2+m^2 r^4 \bigl(\tfrac{2 \lambda}{m} E-\omega^2\bigl)}-\theta \right] \,  ,
\label{eq:trajectunb}
\end{align}
 where we defined the function 
 $$
 \zeta_{[\lambda]}(E)\doteq \frac{2 \bigl(\frac{2 \lambda}{m} E-\omega^2 \bigl)^{\frac{3}{2}}}{\frac{\lambda}{m} E-\omega^2} \, .
 $$
  Similarly to the case of bounded motion, by defining the quantities 
  $$
  \tilde{a}^2_{[\lambda]}\doteq \frac{E}{m(\frac{2 \lambda E}{m}-\omega^2)} \, , \qquad \tilde{\epsilon}_{[\lambda]} \doteq \sqrt{1+\frac{ (\frac{2\lambda E}{m}-\omega^2) l^2}{E^2}} \, ,
  $$
  the expression \eqref{eq:trajectunb} can be written more elegantly as:
\begin{equation}
\zeta_{[\lambda]}(E) t(r)+\theta=\arccos\!\text{h} \left[\frac{1}{\tilde{\epsilon}_{[\lambda]}}\biggl(1+\biggl(\frac{r}{\tilde{a}_{[\lambda]}}\biggl)^2 \biggl)\right]+\frac{\lambda \tilde{a}^2_{[\lambda]} }{1-\lambda \tilde{a}^2_{[\lambda]}} \sqrt{\biggl[1+\biggl(\frac{r}{\tilde{a}_{[\lambda]}}\biggl)^2 \biggl]^2-\tilde{\epsilon}^2_{[\lambda]}}  \ .
\label{eq:tradef}
\end{equation}

 Equation \eqref{eq:tradef} might be compared with the formula $(15)$ given in \cite{PhysicaD}, having in mind the relation existing  between the Hamiltonian $H$ defined in the present paper and the one defined in the aforementioned reference,  here denoted as $\bar{H}$, calculated for $\kappa=\frac{1}{\lambda}$ and $b_j=0$  in~\cite{PhysicaD}. Such relation is given by $ {\bar{H}}=\lambda H-\frac{m \omega^2}{2  }$. In  \cite{PhysicaD}, the equation of motion has been solved assuming that $\bar{E}:= 2 \bar{H}=2 \lambda E-m \omega^2>0$.

%%%%%%%%%%%%%%%%%%%%%%%%%%%%%%%%%%%%%%%%%%%%%%%%%%%

\section{The deformed $\lambda<0$ case}
\label{sec4}

 In this case the deformation parameter is negative and the Hamiltonian can be written as follows
\begin{equation} 
H(r,p)=T(r,p)+V_{\rm eff}(r)=\frac{ p^2}{2m(1-|\lambda| r^2)}+\frac{ l^2}{2m r^2(1-|\lambda| r^2)}+\frac{m \omega^2 r^2}{2(1-|\lambda| r^2)} \, .
\label{hamneg}
\end{equation}
 It is clear that it presents a singularity at the point $r_s=\frac{1}{\sqrt{|\lambda|}}$.

\begin{figure}[t]
\centering
\includegraphics[width=10cm, height=5.4cm]{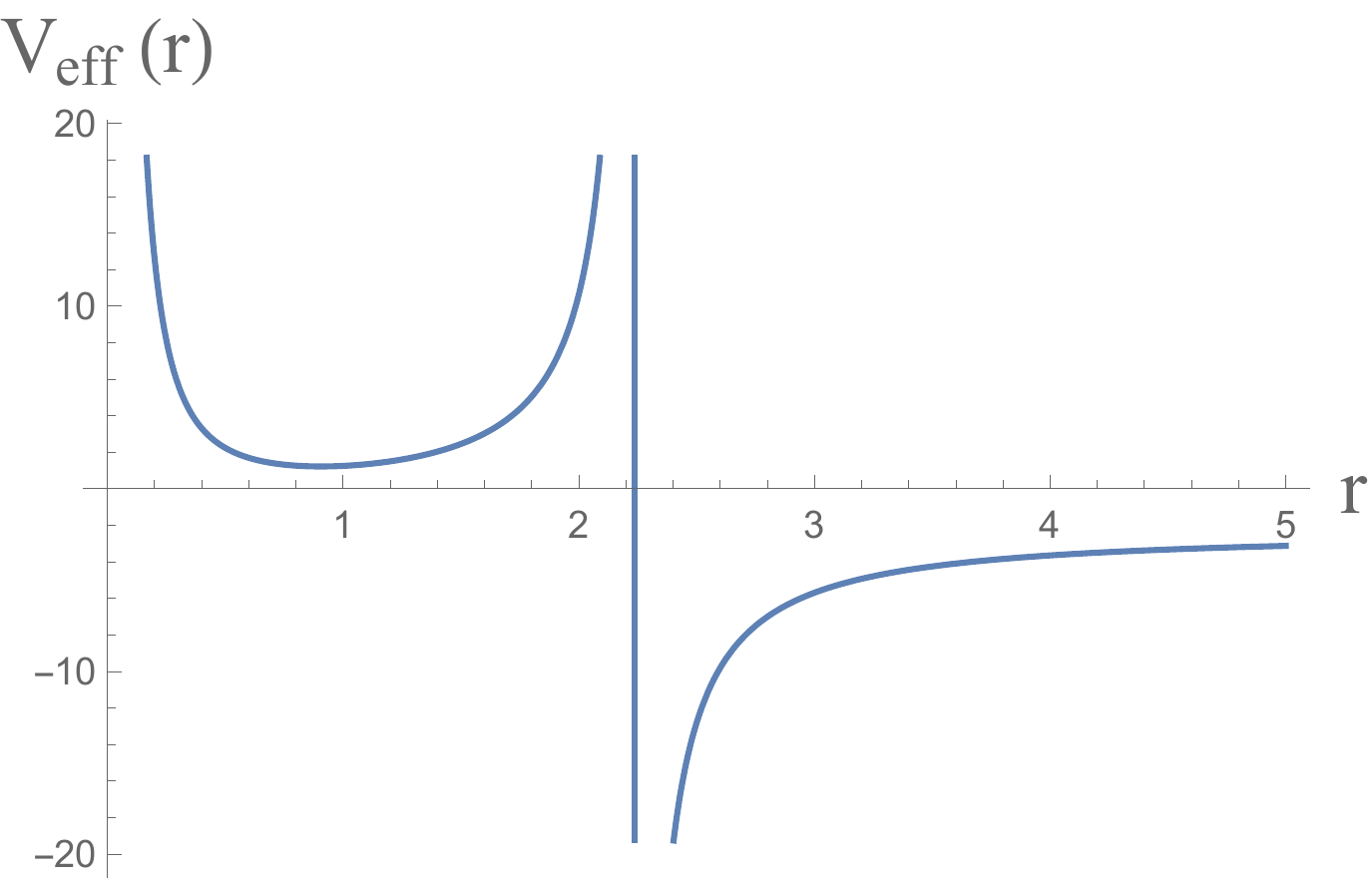}
\caption{Effective potential $V_{\rm eff}(r)$ (\ref{hamneg}) calculated for $m=\omega=l=1$ and $\lambda=-0.20$. The singularity is located at the point $r_s=\sqrt{5}\simeq 2.24$.}
\label{fig7}
\end{figure}

\begin{figure}
\centering 
\includegraphics[width=7.5cm, height=5.4cm]{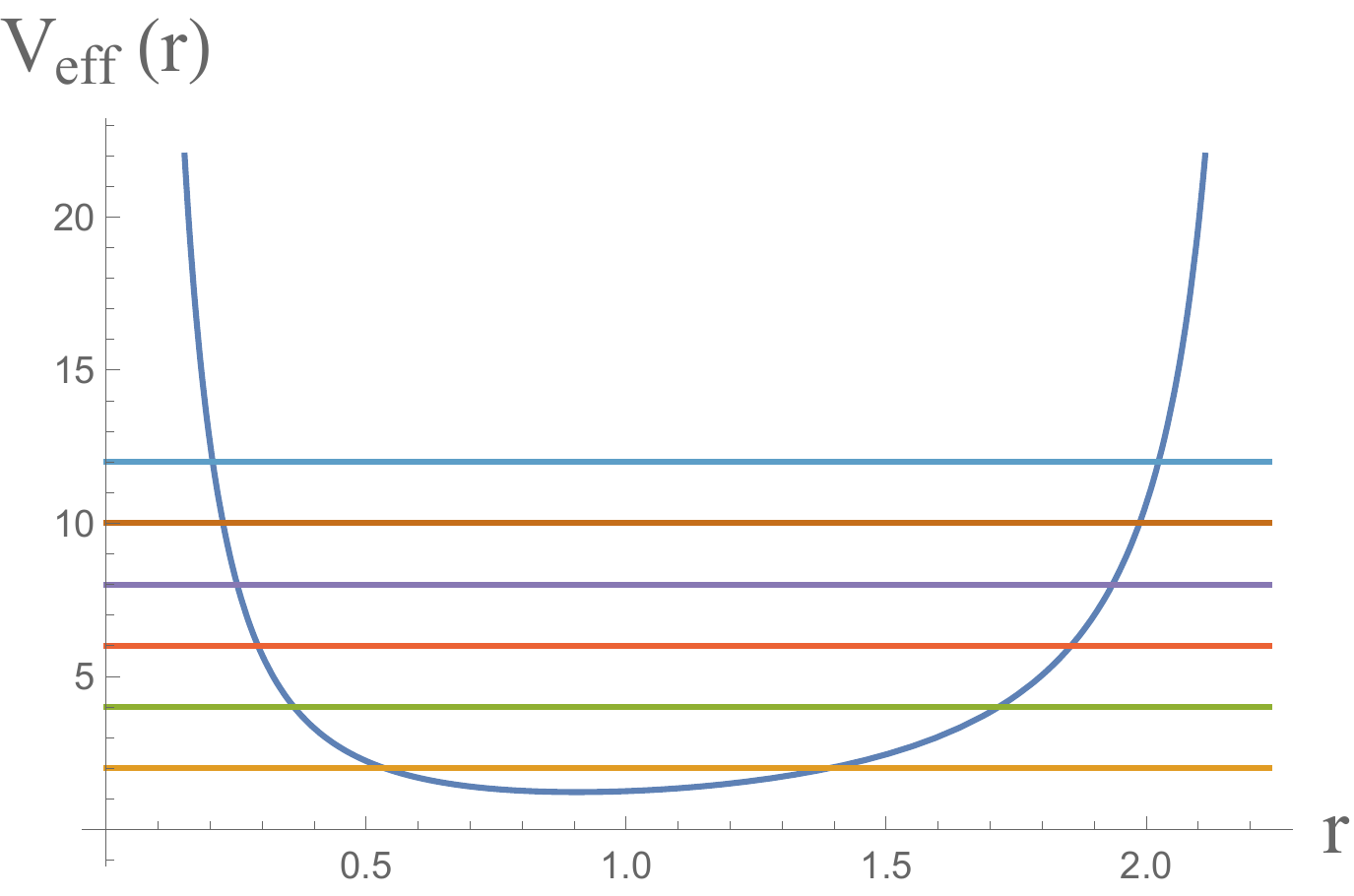}
\quad
\includegraphics[width=7cm, height=5cm]{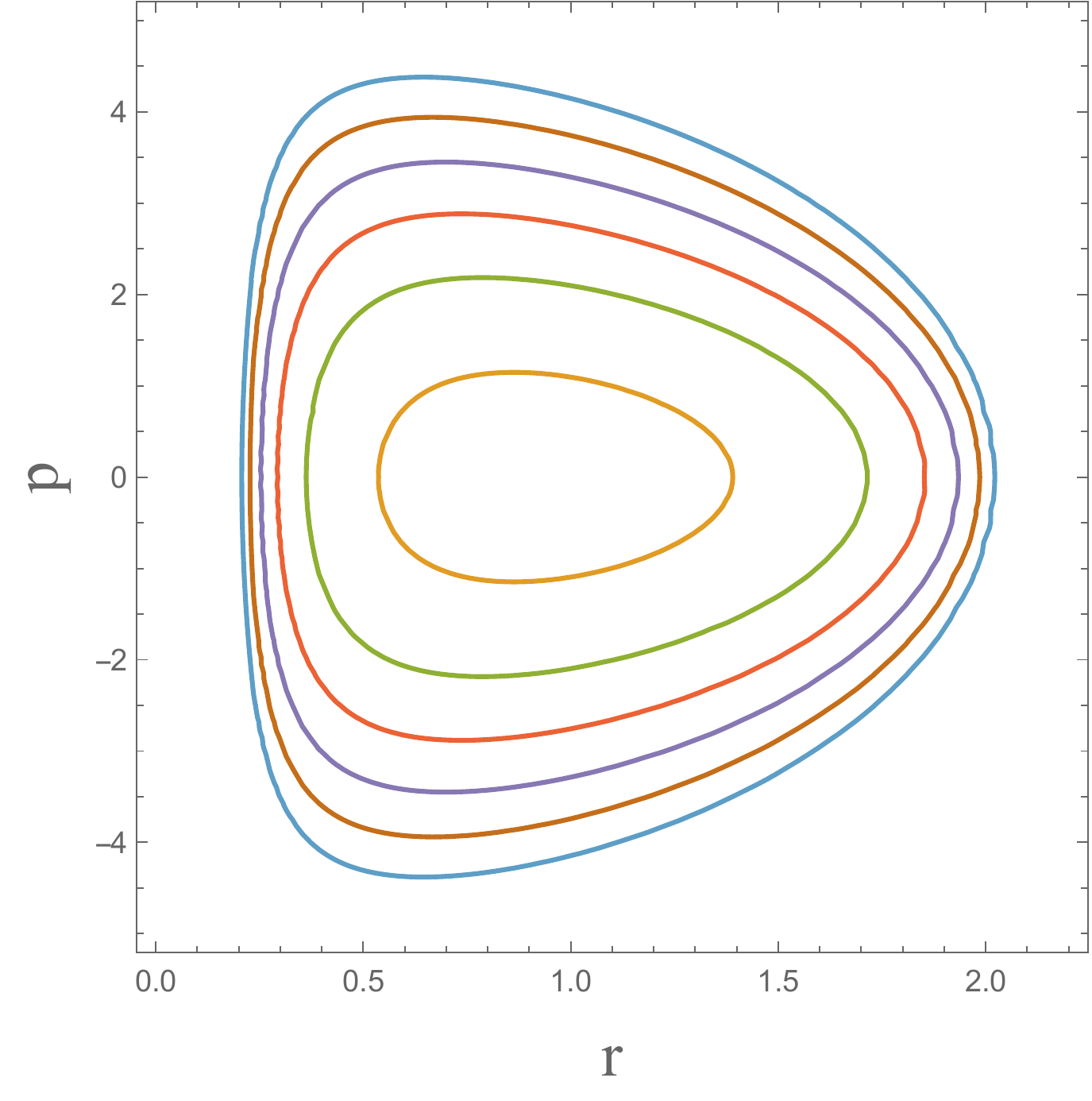}
\caption{Effective potential $V_{\rm eff}(r)$ (\ref{hamneg})  and phase plane $(r, p)$ calculated for $m=\omega=l=1$ and $E=2, 4, 6, 8, 10, 12$ in the region $0<r<r_s$. The deformation parameter is fixed at the value $\lambda=-0.2$.}
\label{fig8}
\end{figure}

 As a matter of fact, owing to the above singularity, the domain of definition of the effective potential splits in two subdomains. The first one, corresponding to  the punctured open ball $0<r<r_s$, is characterized by a positive kinetic energy term in the Hamiltonian and the effective potential exhibits a typical confining shape. Viceversa, the second one, $r_s<r<\infty$, is characterized by a negative kinetic energy and the effective potential has no critical points (see figure \ref{fig7} and lhs of figure \ref{fig8}).

  %%%%%%%%%%%%%%%%%%%%%%%%%%%%%%%%%%%%%%%%%%%%%%%%%%%%
%%%%%%%%%%%%%%%%%%%%%%%%%%%%%%%%%%%%%%%%%%%%%%%%%%%

%%%%%%%%%%%%%%%%%%%%%%%%%%%%%%%%%%%%%%%%%%%%%%%%%%%%%
%%%%%%%%%%%%%%%%%%%%%%%%%%%%%%%%%%%%%%%%%%%%%%%%%%%%%

 The interesting region is the open set $0<r<r_s$. In fact, in this case there will be only closed orbits for energy values $V_{\rm eff}(r_{\rm min})<E<\infty$, where 
 $$
 V_{\rm eff}(r_{\rm min})=\frac{l^2}{m}\left(\sqrt{\lambda^2+\frac{m^2\omega^2}{ l^2}}+|\lambda|\right) \, ,\qquad 
 r^2_{\rm min}=\frac{l^2}{m^2\omega^2}\left(-|\lambda|+\sqrt{\lambda^2+\frac{m^2 \omega^2}{l^2}}\right) \, .
 $$
  In that bounded region the solution can be immediately obtained by means of the SGA technique  by letting in \eqref{eq:tradefo}  $\lambda \to -|\lambda|$. Clearly, since the term  $\omega^2-\frac{2 \lambda}{m} E =\omega^2+\frac{2 |\lambda|}{m} E$ is always positive,  no further restrictions on the energy will apply, i.e.~the energy has no upper bound, as expected. For this case   the phase plane   is depicted in the rhs of figure \ref{fig8}.

  Viceversa, in order to obtain the solution in the unbounded region $r > r_s$, as we already noticed in the case of the Taub-NUT system \cite{CTN}, we have to factorise the new Hamiltonian $\tilde{H}=-H$, the latter  being defined with an overall change of sign. This allows us to restore a positive kinetic energy and then to obtain a physically meaningful Hamiltonian (see also~\cite{IJP2011}). However, this case is not so interesting since no closed orbits are allowed.

%\newpage

%%%%%%%%%%%%%%%%%%%%%%%%%%%%%%%%%%%%%%%%%%%%%%%%%%%
\section{Conclusions}

 To summarize the key points of our paper, we can say that we have successfully applied the SGA approach to the classical D-III oscillator, identifying the structure of its Poisson algebra, and solving the equations of motion  by finding an explicit expression in the form $t=t(r)$. In particular, after having calculated the solution to the equation of motion for the usual isotropic harmonic oscillator (which in our framework just represents the Euclidean limit of the D-III system), we have found the trajectory $t=t(r)$ of the D-III oscillator, for positive values of the deformation parameter $\lambda$,  both in the case of bounded and unbounded motion. Furthermore, the case of  $\lambda$ negative has also been  analyzed. For the latter situation, as we have already seen in the classical Taub-NUT system \cite{CTN},  the most interesting features of the dynamics are somehow hidden `beyond the singularity', in the open set $0<r<r_s$, where for suitable initial conditions the particle remains trapped during its whole life. As a matter of fact, in that region the effective potential shows a typical confining shape and the kinetic energy is always positive definite. We were then enabled to recover an explicit formula for the trajectory, restricted to that domain, simply by taking $t=t(r)$ for $\lambda=-|\lambda|$. 

 In the quantum case we expect similar features, however \emph{in that punctured open ball}  we might miss the algebraic nature of the spectrum  and, accordingly, an analytic expression for the wave functions might not be available.

%%%%%%%%%%%%%%%%%%%%%%%%%%%%%%%%%%%%%%%%%%%%%%%%%%%

\section*{Acknowledgments}

 The research  related to this paper has been developed in the framework of the PRIN prot.~n. $2010$JJ$4$KPA\_$004$. It has   been partially supported by the Spanish  Ministerio de Econom{\'{\i}}a y Competitividad     (MINECO)   under grant MTM2013-43820-P  and by the Spanish Junta de Castilla y Le\'on under grant  BU278U14.

%%%%%%%%%%%%%%%%%%%%%%%%%%%%%%%%%%%%%%%%%%%%%%%%%%%


\begin{thebibliography}{99}

\bibitem{AP2011}
Ballesteros A, Enciso A, Herranz F J, Ragnisco O and Riglioni D $2011$ \emph{Ann. Phys.} {\bf 326} 2053

 


 \bibitem{Konig}
 Koenigs G 1972   {\em    {{L}e{\c{c}}ons sur la th\`eorie g\`en\`erale des surfaces}}
  vol  4  ed   G Darboux (New York: Chelsea) p  368
  
\bibitem{JMP2003}
 Kalnins E G,  Kress J M, Miller W Jr and Winternitz P 2003
 {\em J. Math. Phys.} {\bf 44}    5811


\bibitem{Pogosyan} Grosche C,  Pogosyan G S and Sissakian A N  2007
{\em Phys. Part.   Nuclei} {\bf  38}  525 



\bibitem{PLB}
Ballesteros A, Enciso A, Herranz F J and  Ragnisco O $2007$ \emph{Phys. Lett. B} {\bf 652}  376


\bibitem{PhysicaD}
 Ballesteros A, Enciso A, Herranz F J and Ragnisco O  $2008$ \emph{Physica D} {\bf 237} 505


 \bibitem{AP2009}
Ballesteros A, Enciso A, Herranz F J and  Ragnisco O $2009$  \emph{Ann. Phys.} {\bf 324} 1219


\bibitem{Bertrand}
Bertrand J $1873$ \emph{C.R. Math. Acad. Sci. Paris} {\bf 77} 849

\bibitem{Perlick}
Perlick V $1992$ \emph{Class. Quantum Grav.} {\bf 9} 1009

\bibitem{CQG2008}
Ballesteros A, Enciso A, Herranz F J and Ragnisco O $2008$ \emph{Class. Quantum Grav.} {\bf 25} 165005

\bibitem{CMP2009}
Ballesteros A, Enciso A,  Herranz F J and Ragnisco O $2009$ \emph{Commun. Math. Phys.} {\bf 290} 1033



\bibitem{PLA}
Ballesteros A, Enciso A,  Herranz F J,  Ragnisco O and    Riglioni D $2011$ \emph{Phys. Lett. A} {\bf 375} 1431


\bibitem{IJP2011} 
Ballesteros A, Enciso A,  Herranz F J,  Ragnisco O and    Riglioni D $2011$  \emph{Int. J. Theor. Phys.} {\bf 50} 2268





 \bibitem{Demkov}
  Demkov Y N 1959  {\em Soviet Phys. JETP} {\bf 9} 63 


 \bibitem{AJ1965}
Fradkin D M 1965     {\em Amer. J. Phys.} {\bf 33}  207 





\bibitem {AP2008} 
Kuru S and Negro J $2008$ \emph{Ann. Phys.} {\bf 323}  413

\bibitem{JPCS2012} 
Kuru S and  Negro J $2012$ \emph{J. Phys.: Conf. Ser.} {\bf 343} 012063

\bibitem{CTN}
Latini D and Ragnisco O $2015$ \emph{J. Phys. A: Math. Teor.}  {\bf 48} 175201


\end{thebibliography}
\end{document}